\documentclass[aps,amsmath,amssymb,pra,reprint]{revtex4-2}

\usepackage[english]{babel}
\usepackage{graphicx}
\usepackage{bm}
\usepackage[usenames]{xcolor}
\usepackage{mathrsfs,verbatim}
\usepackage[normalem]{ulem}

\usepackage[plainpages=false,
pdfpagelabels,
colorlinks=true,linkcolor=red,urlcolor=blue,citecolor=blue,
pdftitle={Universal Landau-Zener regimes in dynamical topological phase transitions},
pdfauthor={Yang Ge, Marcos Rigol},
pdfdisplaydoctitle=true]{hyperref}

\newcommand{\mathe}{\mathrm{e}}
\newcommand{\mathi}{\mathrm{i}}

\newcommand{\mathpi}{\pi}

\newcommand{\tmop}[1]{\ensuremath{\operatorname{#1}}}

\newcommand{\ket}[1]{\ensuremath{| #1 \rangle}}
\newcommand{\bra}[1]{\ensuremath{\langle #1 |}}
\newcommand{\braket}[2]{\ensuremath{\langle #1 | #2 \rangle}}
\newcommand{\assign}{\ensuremath{:=}}

\begin{document}

\title{Universal Landau-Zener regimes in dynamical topological phase transitions}

\author{Yang Ge}

\author{Marcos Rigol}

\affiliation{Department of Physics, The Pennsylvania State University, University Park, Pennsylvania 16802, USA}

\begin{abstract}
  In finite systems driven unitarily across topological phase transitions, the Chern number and the Bott index have been found to exhibit different behaviors depending on the boundary conditions and on the commensurability of the lattice. For periodic boundary conditions, the Chern number does not change for finite commensurate lattices (or in the thermodynamic limit). On the other hand, the Chern number can change for incommensurate lattices with periodic boundary conditions and the Bott index can change for lattices with open boundary conditions. Here we show that the scalings of the fields at which those two indices change exhibit Landau-Zener and near-adiabatic regimes depending on the speed at which the strength of the drive is ramped up and on the system size. Those regimes are preceded by a regime in which the topological indices do not change. The latter is the only regime that, for nonvanishing ramp speeds, survives in the thermodynamic limit. We then show that the dc Hall response can be used to detect topological phase transitions independently of the behavior of the topological indices.
\end{abstract}

{\maketitle}

\section{Introduction}\label{sec:introduction}

Dynamically generating topological phases of matter is a powerful way to explore a variety of topological states and theoretical models that cannot or are very challenging to explore in equilibrium~\cite{jotzu_experimental_2014, bandres_topological_2018, zilberberg_photonic_2018, cooper_topological_2019, zhang_topological_qm_cold_atoms_2019, mciver_light-induced_2020}. Periodic driving, e.g., shining circularly polarized light on a material, is one of the most promising protocols that has been proposed to dynamically generate (Floquet) topological states~\cite{oka_photovoltaic_2009, inoue_photoinduced_2010, kitagawa_transport_2011, lindner_floquet_2011, bauer_topologically_2019, oka_floquet_2019, ozawa_topological_2019}. A common feature of most of those protocols is that the Floquet topological states of interest are not adiabatically connected to the initial (trivial) equilibrium states of the systems that are periodically driven. This has generated much interest in what happens in real time when systems are driven across topological phase transitions \cite{heyl_dynamical_2018, foster_quantum_2013, foster_quench-induced_2014, tarnowski_observation_2017, flaschner_observation_2018, tarnowski_measuring_2019, tian_observation_2019, zhang_dynamical_2018, yi_observing_2019, zhang_dynamical_2019, hu_topological_2020, unal_topological_2020}.

For translationally invariant systems in the thermodynamic limit, it was proved that under driven unitary dynamics the Chern number does not change when crossing a topological phase transition~\cite{dalessio_dynamical_2015, caio_hall_2016}. More generally, the invariance of the Chern number (among other topological indices) is ensured by the fixed-point theorem in topology and the continuity of dynamics. The  fixed-point theorem in topology implies that phases with distinct Chern numbers have orthogonal many-body wavefunctions (e.g., in a two-band system there is always a momentum point at which states of different Chern numbers point in antipodal directions on the Bloch sphere), while in continuous dynamics it is impossible for $\langle \psi(t + d t) | \psi(t) \rangle = 0$ when the energy of the state is bounded. Hence, naive formulations of topological indices are generally invariant under unitary time evolution in the thermodynamic limit.

The experimental observations of a nonzero Hall drift~\cite{jotzu_experimental_2014, aidelsburger_measuring_2015} thus inspired a plethora of studies~\cite{foster_quantum_2013, dehghani_dissipative_2014, sacramento_fate_2014, dalessio_dynamical_2015, dehghani_out_2015, sacramento_charge_2015, dehghani_occupation_2016, sacramento_edge_2016, hu_dynamical_2016, privitera_quantum_2016, caio_quantum_2015, caio_hall_2016, ge_topological_2017, caio_topological_2019, ulcakar_slow_2018, ulcakar_slow_2019, ulcakar_kibble-zurek_2020, bandyopadhyay_unitary_2020}. Nonunitary effects, including thermal baths~\cite{dehghani_dissipative_2014, dehghani_out_2015} and dephasing noise~\cite{hu_dynamical_2016}, were shown to stabilize a nearly quantized Hall conductance, in much the same way that a diagonal ensemble of unitary time evolution would predict~\cite{wang_phase_2016, wang_universal_2016}. For quantum quenches, another protocol used to generate topological states~\cite{foster_quench-induced_2014}, it was found that two different formulations of the bulk topological winding number that are equivalent for topological superconductors in equilibrium are nonequivalent out of equilibrium (one being generically conserved and the other not)~\cite{foster_quantum_2013}.

Another question, the focus of this work, is the role of the boundary conditions in the dynamics of topological indices. For systems with open boundary conditions, or in general for systems that are not translationally invariant, the Bott index is the topological index that is commonly used to characterize topological phases in equilibrium~\cite{loring_disordered_2010, hastings_topological_2011, titum_disorder-induced_2015}. When computed out of equilibrium in systems with open boundary conditions driven across a topological phase transition, the Bott index can change~\cite{dalessio_dynamical_2015, toniolo_time-dependent_2018}. (The local Chern marker~\cite{bianco_mapping_2011} was found to exhibit similar behavior in Ref.~\cite{privitera_quantum_2016}.) One can also compute the Bott index in finite translationally invariant systems~\cite{ge_topological_2017}, in which it is equivalent to the discretized Chern number usually used in finite-system calculations~\cite{fukui_chern_2005}. In Ref.~\cite{ge_topological_2017}, we showed that the discretized Chern number can change in finite translationally invariant systems that are incommensurate, namely, in translationally invariant systems in which the gap-closing point(s) is(are) missing in their Brillouin zone. On the other hand, the discretized Chern number does not change in commensurate systems, namely, those that contain the gap-closing point(s) in their Brillouin zone.

One of the goals of this work is to provide a unified picture of the real-time dynamics of topological indices in finite periodically driven systems with periodic and open boundary conditions. We also discuss the scaling with system size of the critical field of topological transitions in the Floquet Hamiltonian in lattices with periodic and open boundary conditions. Another goal is to show that the dc Hall response can be used to detect topological phase transitions independently of the behavior of the discretized Chern number or the Bott index. 

The presentation is organized as follows. In Sec.~\ref{sec:setup}, we discuss the model, the drive protocol used to generate Floquet topological phases, and the geometries considered. The finite-size scaling of the critical field of the topological transition of interest in the Floquet Hamiltonian is discussed in Sec.~\ref{sec:equilibrium}. In that section, we review results obtained in Ref.~\cite{ge_topological_2017} for periodic boundary conditions and report results for patch and cylinder geometries. In Sec.~\ref{sec:dynamics} we study the scaling of the critical field of the topological transition during the unitary dynamics generated by ramping up the driving field. We consider incommensurate translationally invariant systems as well as patch and cylinder geometries. Section~\ref{sec:hall} is devoted to the study of the dc Hall response in the topologically trivial and nontrivial ground states of the Floquet Hamiltonian in translationally invariant systems, as well as in the nonequilibrium state generated after ramping up the driving field. A summary and discussion of our results are presented in Sec.~\ref{sec:summary}.

\section{\label{sec:setup}Model, drive protocol, and geometries}

\begin{figure}[!t]
  \includegraphics[width=\columnwidth]{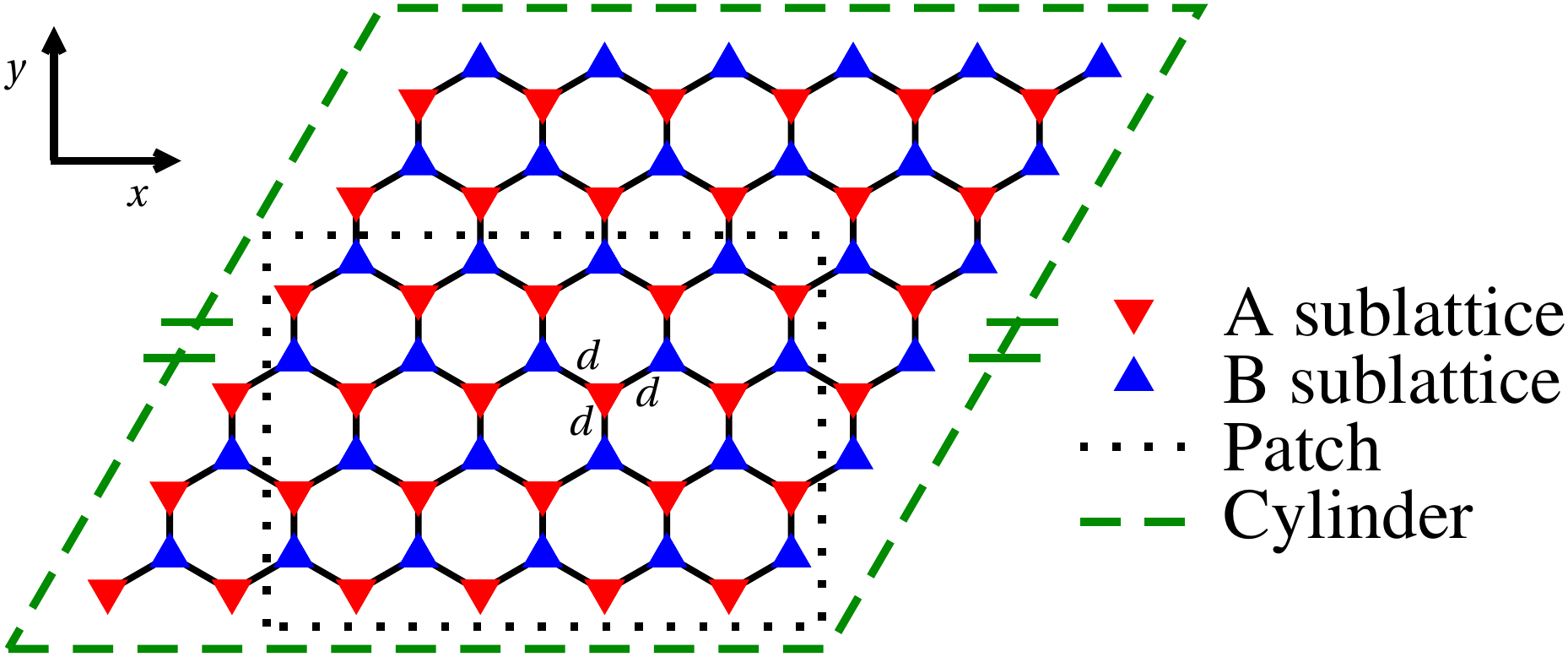}
  \caption{\label{fig:lat}
    Honeycomb lattice with nearest-neighbor lattice spacing $d$, and highlighted sublattices ${\mathcal A}$ and ${\mathcal B}$. The dotted line indicates the boundaries of the patch geometry (open boundaries in both directions), while the dashed line indicates the boundary of the cylinder geometry (translationally invariant in the horizontal direction and zigzag open boundaries in the vertical one). In our calculations, all cylinders have the same number of unit cells along both lattice directions, while lattices in the patch geometry have a roughly equal number of unit cells in the $x$ and $y$ directions.}
\end{figure}

We study spinless fermions on a honeycomb lattice with nearest-neighbor hopping and a staggered potential between the $\mathcal{A}$ and $\mathcal{B}$ sublattice sites, as displayed in Fig.~\ref{fig:lat}. An in-plane circularly polarized oscillating electric field is used as the Floquet drive. Setting $\hbar = 1$, the Hamiltonian reads
\begin{equation}
  \label{eq:hamiltonian} \hat{H} (t) = - J \sum_{\langle i, j \rangle} \left[ \mathe^{\mathi e \vec{A} (t) \vec{d}_{i j}}  \hat{c}_i^{\dag} \hat{c}_j + \text{H.c.} \right] + \frac{\Delta}{2} \sum_{\substack{ i \in \mathcal{A}\\ j \in \mathcal{B} }} (\hat{n}_i - \hat{n}_j),
\end{equation}
where the vector potential $\vec{A} (t) = A (\sin \Omega t, \cos \Omega t)$ describes the drive's electric field with frequency $\Omega$ (we label the period of the drive as $T=2\pi/\Omega$). We set the strength of the staggered potential $\Delta = 0.15 J$ to match the experiment in Ref.~\cite{jotzu_experimental_2014}. This value of the staggered potential opens a gap in the static system generating a trivial insulator at half filling.

Depending on the frequency and strength of the drive, the resulting Floquet Hamiltonian $\hat H_F$, defined from
\begin{equation}
    \hat U(T)=\exp[-i\hat H_F T],
\end{equation}
where $\hat U(T)$ is the evolution operator over one period,
displays various topological phases at half filling~\cite{mikami_brillouin-wigner_2016, ge_topological_2017}. Here we focus on the first transition in the high-frequency regime, for $\Omega = 7 J$ (this value of $\Omega$ is greater than the band width). In this regime, to the lowest order in the drive period~\cite{ge_topological_2017, mikami_brillouin-wigner_2016}, the Floquet Hamiltonian is the Haldane model~\cite{haldane_model_1988}. The topological transition in the thermodynamic limit occurs at $edA^{\ast} \approx 0.498$. In what follows, we set $e = d = 1$.

To dynamically prepare a topological state starting from the trivial band insulator at $A=0$, we turn on the electric field using a linear ramp with ramp time $\tau$, such that
\begin{equation} \label{eq:ramp-path}
    A (t) = t / \tau .
\end{equation}
Different ramping speeds, system sizes, and, in the translationally invariant case, the commensurability of the lattice result in different behaviors of the topological indices as $A (t)$ crosses the critical field.

We study finite lattices with three different boundary conditions (see Fig.~\ref{fig:lat}): (i) periodic (tori) so that the systems are translationally invariant, (ii) periodic in one direction and open in the other one (cylinders), and (iii) open in both directions (patches). To each of those lattice geometries we associate a specific topological index that allows one to locate the topological transition in the Floquet Hamiltonian. For periodic boundary conditions we use the discretized Chern number (in short, the Chern number)~\cite{fukui_chern_2005, ge_topological_2017}, while for the patch geometry we use the Bott index~\cite{loring_disordered_2010, dalessio_long-time_2014}. 

For the cylinder geometry, Fourier expanding in momentum space along the translational invariant direction $x$, the Bott index can be rewritten using the projection operator $\hat{P}_{k_x}$ onto the periodic part of Bloch states within filled bands at each $k_x$ (see Appendix~\ref{sec:cylinder-bott}):
\begin{equation}
\label{eq:bott-cylinder}
  C_b (\hat{P}) = \frac{1}{2 \mathpi} \tmop{Im} \sum_{k_x} \tmop{Tr} \ln ( \hat{P}_{k_x} \mathe ^{\mathi \delta_y  \hat{y}} \hat{P}_{k_x}  \hat{P}_{k_{x-}} \mathe ^{-\mathi \delta_y  \hat{y}} \hat{P}_{k_{x-}} ),
\end{equation}
where $\delta_{y} = 2 \mathpi / L_{y}$, $k_{x-} = k_x-2\mathpi / L_x$. The logarithm and the trace should be evaluated in the subspace spanned by filled bands at each $k_x$. This formula is a discretized integral that calculates the Wannier center flow winding along $x$ {\cite{asboth_short_2016}}. A technical complication arises because gapless edge states make Fermi seas ill defined in topologically nontrivial phases. It is then necessary to include in the calculation the relevant (nearly) degenerate states at the $k$ points at which the edge modes cross, and then discard the kernel of the operator product due to the extra state before taking the logarithm. Our dynamical calculations are not affected by this complication because the initial state is a topologically trivial insulating state. The above formulations of the Chern number and the Bott index are equivalent in the thermodynamic limit.

\section{\label{sec:equilibrium}Scaling of the critical field in the Floquet Hamiltonian}

In finite lattices, the critical fields $A^{\ast}_L$ at which topological transitions occur in the Floquet Hamiltonian (the fields at which the topological indices change value) depend on the linear system size $L$. For translationally invariant systems, the size dependence of the critical field for the topological transition of interest here was studied in detail in Ref.~\cite{ge_topological_2017}. There we showed that the commensurability of the lattice plays an essential role. In commensurate lattices, with linear dimensions $L=3n$ ($n\in \mathbb{Z}$) which contain the $K$ and $K'$ points  [$(\frac{2 \pi}{3},\frac{4 \pi}{3})$ and $(\frac{4\pi}{3},\frac{2\pi}{3})$, respectively] at which gaps may close at the topological transitions, the critical field is system-size independent starting from $L=3$. On the other hand, in lattices with $L=3n+1$ and $L=3n+2$ the critical field approaches the thermodynamic limit result as $\sim 1/L^2$. This occurs because in finite incommensurate lattices the Chern number cannot resolve the singular behavior of Berry curvature at the phase transition until it spreads out in momentum space as the systems go deeper into the topological regime.

\begin{figure}[!b]
  \includegraphics[width=\columnwidth]{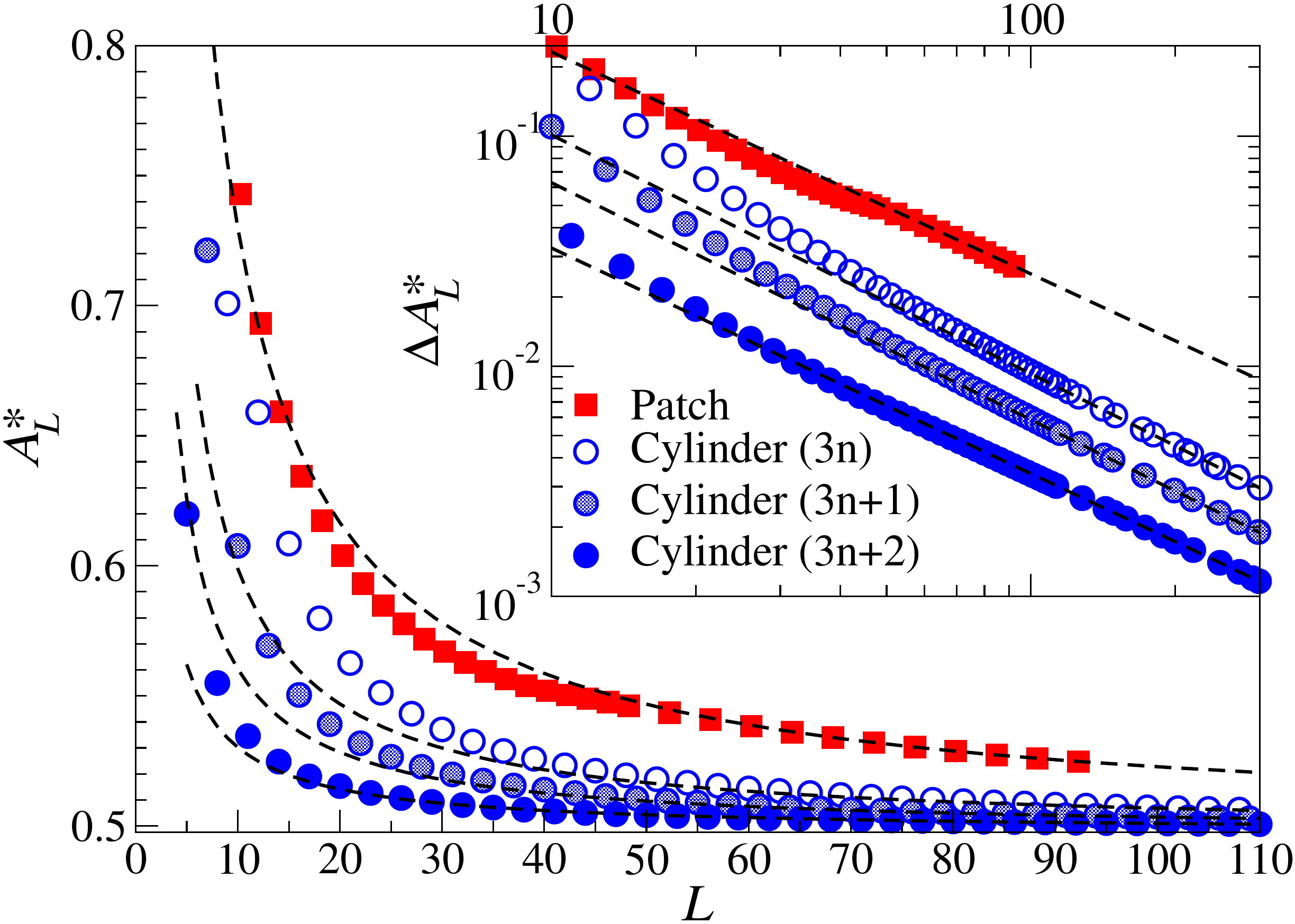}
  \caption{\label{fig:eq-trans}
    Critical strength of the drive's field $A^{\ast}_L$ plotted as a function of $L$ for cylinder and patch geometries. For the cylinders $L$ is the number of unit cells in each lattice direction, while for the patch geometry $L$ is the square root of the total number of unit cells (we keep $L_x\simeq L_y$). The inset shows the offset $\Delta A^{\ast}_L = A^{\ast}_L - A^{\ast}$ from the critical field strength $A^{\ast}$ in the thermodynamic limit. We fit power laws $\Delta A^{\ast}_L \propto L^{- 2 \alpha}$ to the largest system sizes depicted, obtaining $\alpha \approx 0.483$ for the patch geometry, 0.519 for cylinders with $L=3n$, 0.515 for cylinders with $L=3n+1$, and 0.491 for cylinders with $L=3n+2$. The dashed lines in the main panel and inset depict the results of our fits. Note that, in cylinders, finite-size effects are smallest for $L=3n+2$. These are the cylinders on which we focus in the remainder of this work.}
\end{figure}

In this section, we use the Bott index (see Sec.~\ref{sec:setup}) to determine $A^{\ast}_L$ for the Floquet Hamiltonian in cylinder and patch geometries, and to study the system-size dependence of $A^{\ast}_L$. Figure~\ref{fig:eq-trans} shows the results obtained for $A^{\ast}_L$ vs $L$. For the cylinder geometry $L$ is the (equal) number of unit cells in each lattice direction, while for the patch geometry $L$ is defined as the square root of the total number of unit cells and we keep roughly an equal number of unit cells in the $x$ and $y$ directions. While $A^{\ast}_L$ depends smoothly on $L$ for the patch geometry, we find strong commensurability effects in the cylinder geometry. In the latter, a smooth dependence on $L$ is seen only if one looks independently at the results for $L = 3 n$, $3 n+1$, and $3 n+2$ (finite-size effects being strongest for $L = 3 n$ and weakest for $L = 3 n+2$). Still, for both geometries one can see that $A^{\ast}_L$ approaches the thermodynamic limit result $A^{\ast} \approx 0.498$ with increasing system size. 

In the inset in Fig.~\ref{fig:eq-trans} we plot $\Delta A^{\ast}_L = A^{\ast}_L - A^{\ast}$ vs $L$. These plots make apparent that $A^{\ast}_L$ approaches the thermodynamic limit result as a power law in $L$. Power-law fits of $\Delta A^{\ast}_L$ for the largest systems sizes (depicted as dashed lines in the main panel and inset in Fig.~\ref{fig:eq-trans}) reveal that $\Delta A^{\ast}_L \sim 1/L$. (The local Chern marker exhibits similar scalings in systems with open boundary conditions~\cite{caio_topological_2019}.) Our scalings in Fig.~\ref{fig:eq-trans} are to be contrasted to $\Delta A^{\ast}_L\sim 1/L^2$ found for incommensurate lattices with periodic boundary conditions. Since cylinders with $L = 3n+2$ exhibit the smallest finite-size effects, those will be the ones on which we focus in the reminder of this work.

\section{\label{sec:dynamics}Scaling of the critical field in the dynamics}

In the thermodynamic limit, and in finite translationally invariant lattices that are commensurate, both the Chern number and the Bott index of fully filled bands are invariant upon unitarily driving the system~\cite{dalessio_dynamical_2015, caio_hall_2016, ge_topological_2017, toniolo_time-dependent_2018}. This comes from the unavoidable diabatic bulk excitation at the phase transition, which occurs because of the closing of a gap. Such a gap closing is inevitable in the transition studied here since all symmetries are already broken. Hence, no detour via symmetry breaking exists \cite{mcginley_classification_2019}, unlike, for example, the case of the Su-Schrieffer-Heeger model \cite{song_observation_2018, bandyopadhyay_dynamical_2019}. 

In this section we study the dynamical regimes that arise in translationally invariant lattices that are incommensurate, as well as in patch and cylinder geometries, depending on the system size and the speed at which one ramps up the strength of the drive's field.

\subsection{Translationally invariant systems}

In Ref.~\cite{ge_topological_2017} we showed that, in translationally invariant two-band systems [as described by Eq.~\eqref{eq:hamiltonian}], unitarily driving across a topological transition generates Landau-Zener transitions \cite{privitera_quantum_2016, hu_dynamical_2016, breuer_quantum_1989}. In the topological transition explored there (the same of interest here), the gap closes at the $K'$ Dirac point so the momentum states considered were near that point. In Ref.~\cite{ge_topological_2017}, we also showed that the Chern number can change in incommensurate systems for slow enough ramps. Now we connect those results showing that in incommensurate translationally invariant systems the finite-size scaling of the excitations and of the critical field when crossing the topological transition are described by the Landau-Zener formulas. 

The Landau-Zener theory describes coupled two-state systems in which the diagonal (uncoupled) energies change linearly in time~\cite{zener_non-adiabatic_1932, landau_landau-zener_1932, wittig_landauzener_2005}, 
\begin{equation}
  H(t) = \left( \begin{array}{cc} \beta t & V\\  V & - \beta t \end{array} \right),
\end{equation}
with the system initially in the ground state at $t=-\infty$. A finite coupling $V$ allows transitions to occur, and at $t=\infty$ the probability of finding the system in the excited state is
\begin{equation}
  \label{eq:LZ-excitation}
  P_E = \exp(-\pi V^2 / \beta).
\end{equation}
For our topological transition, the Landau-Zener formula gives the final occupation in the higher Floquet band. For incommensurate translationally invariant systems, $V \sim 1 / L$ and $\beta \sim 1 / \tau$, hence $V^2 / \beta \sim \tau / L^2$. Henceforth, we define $\tau / (L^2 T)$ to be the dimensionless Landau-Zener parameter (in short, the Landau-Zener parameter). Furthermore, we evaluate Eq.~\eqref{eq:LZ-excitation} via determining the exact relationships between $V$ and $L$, as well as $\beta$ and $\tau$, using: (i) that $V = \eta v_F / L$ at the gap-closing point at $A^\ast$ ($v_F$ is the Fermi velocity), with $\eta=2/9$ for our transition (the gap closes at the $K'$ point), and (ii) that $\beta$ is the rate of increase of the time-reversal-breaking mass term generated by the Floquet drive near criticality~\cite{titum_disorder-induced_2015, haldane_model_1988}.  

\begin{figure}[!t]
  \includegraphics[width=\columnwidth]{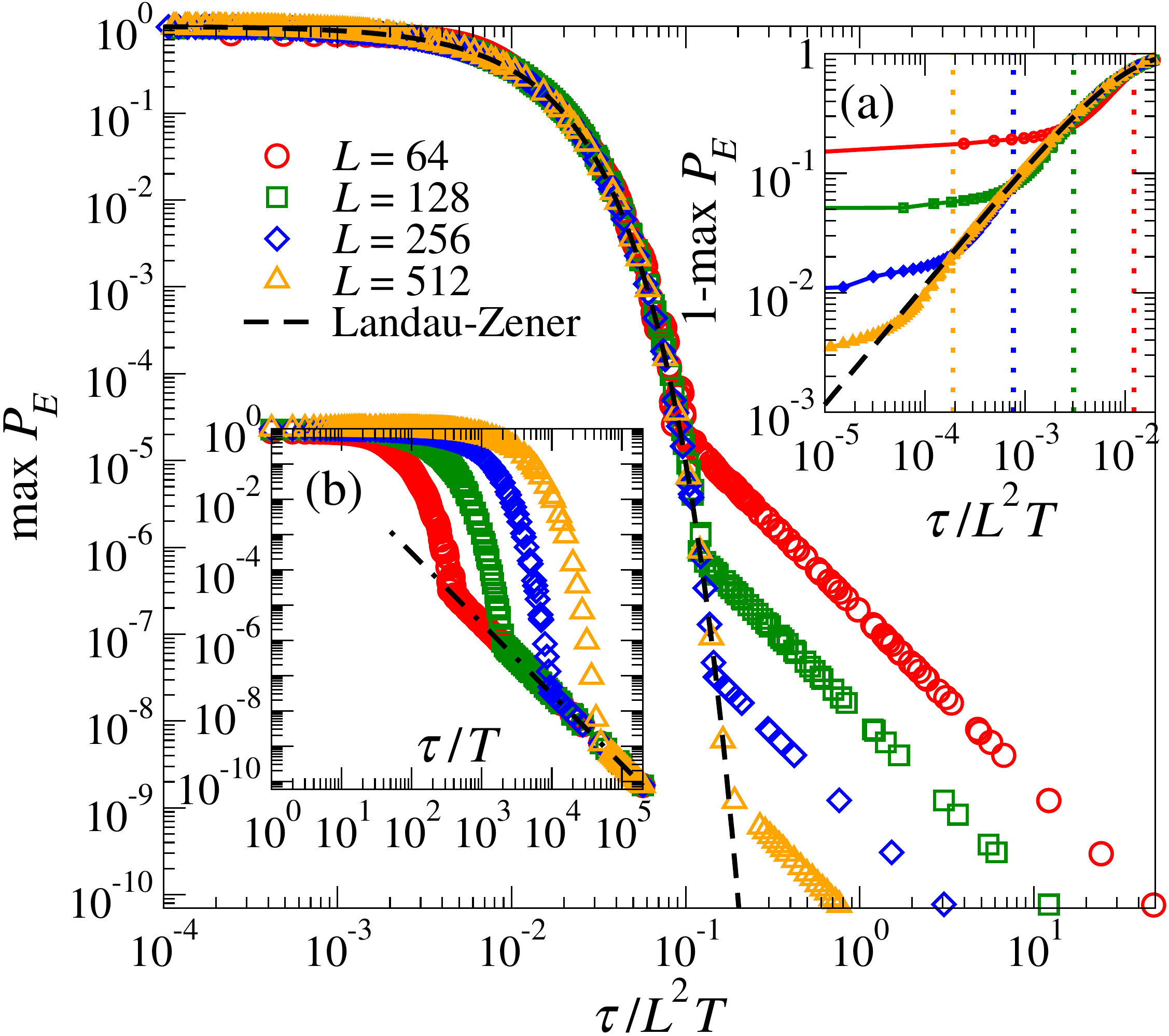}
  \caption{\label{fig:lz-overlap}
    Highest occupation of any $k$ point in the conduction Floquet band across the Brillouin zone, in incommensurate lattices with $L = 3 n+1$ and $3 n+2$, plotted as a function of the Landau-Zener parameter. We show results for four lattice sizes. The dashed line is the Landau-Zener prediction [Eq.~\eqref{eq:LZ-excitation}]. (a) The departure of the maximum occupation from 1 in the limit of fast ramps. The dashed line is the Landau-Zener prediction, while the vertical dotted lines mark the $\tau=50T$ boundary between the fast-ramp and the Landau-Zener regimes identified in Fig.~\ref{fig:lz-overlap-full}(b). (b) The maximum excitation plotted as a function of the ramp time $\tau$, highlighting the near-adiabatic regime in which $P_E \sim \tau^{-2}$ (the dash-dotted line depicts $\tau^{-2}$ behavior).} 
\end{figure}

While translationally invariant commensurate systems are fully excited at the gap-closing momentum after the topological transition is crossed ($V=0$), this momentum point is absent in incommensurate systems. Figure~\ref{fig:lz-overlap} shows the highest occupation of any $k$ point in the conduction Floquet band (dubbed the maximum excitation) across the Brillouin zone, in incommensurate lattices with $L = 3 n+1$ and $3 n+2$, plotted as a function of the Landau-Zener parameter. The dashed line shows the Landau-Zener prediction [Eq.~\eqref{eq:LZ-excitation}]. Three regimes can be identified: 

(i) For fast ramps, the maximum excitation (very close to 1) depends weakly on $\tau$ [see Fig.~\ref{fig:lz-overlap}(a)]. The maximum excitation in this regime is essentially determined by the overlap between the initial and final valence-band states, which depends on the system size. 

(ii) For intermediate ramp speeds, one can see a collapse of the maximum excitation for different lattice sizes when plotted against the Landau-Zener parameter. This is a regime in which Landau-Zener dynamics dominates and the maximal excitation occurs in a $k$ point close to the gap-closing momentum. 

(iii) For slow ramps, the system follows near adiabatically the drive. The maximum excitation no longer exhibits collapse when plotted against the Landau-Zener parameter. In this regime, diabatic processes other than those about the gap-closing momentum, or those about the critical field strength, may dominate the Fermi sea excitation. Also, the maximum excitation is solely determined by the ramping speed $P_E \sim \tau^{-2}$ as expected~\cite{WEINBERG20171} [see Fig.~\ref{fig:lz-overlap}(b)]. 

\begin{figure}[!t]
  \includegraphics[width=\columnwidth]{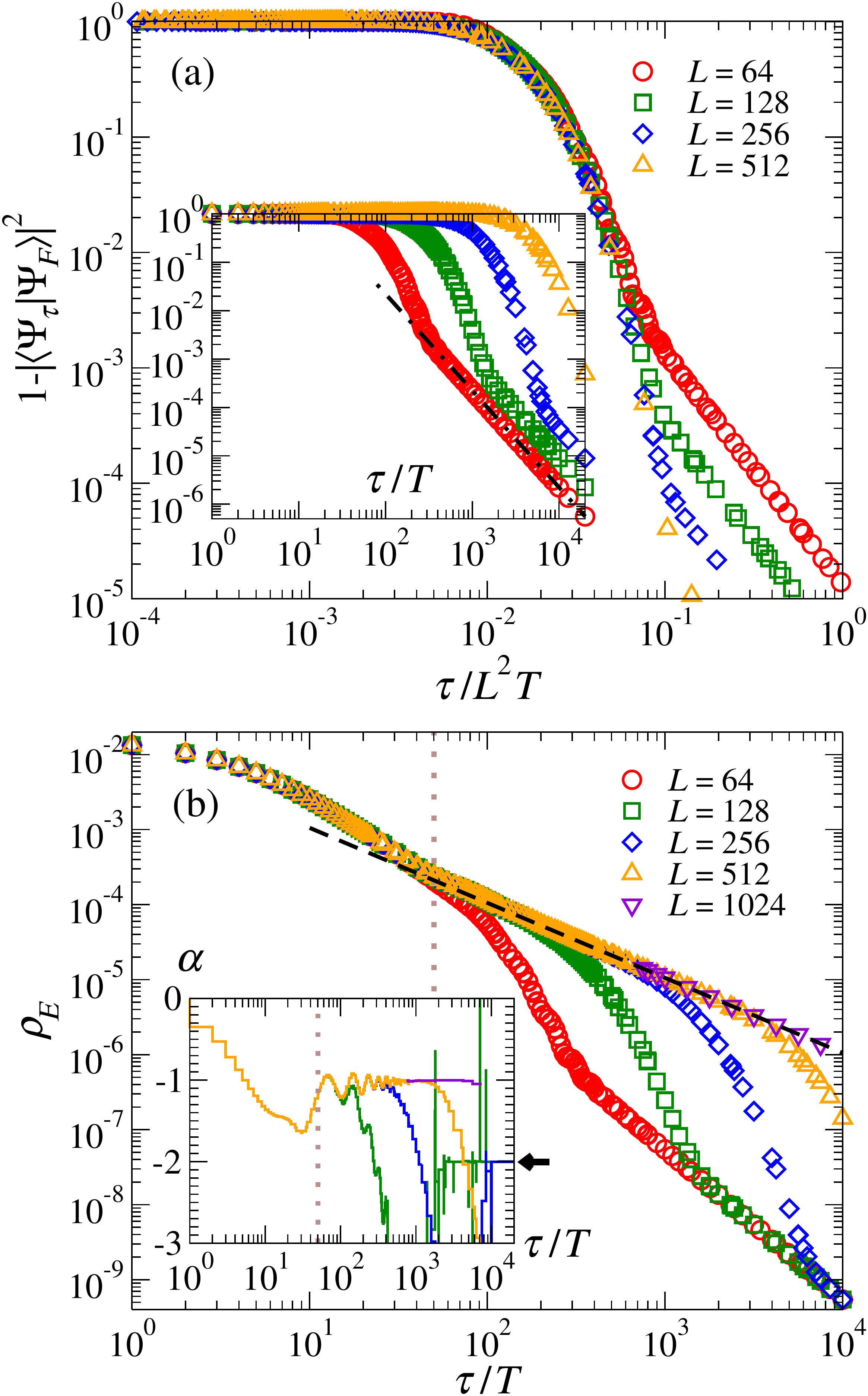}
  \caption{\label{fig:lz-overlap-full}
    (a) $ 1 - | \langle \Psi_\tau | \Psi_{F} \rangle |^2 $, where $\ket{\Psi_\tau}$ is the time-evolved state and $| \Psi_{F} \rangle$ is the ground state of the Floquet Hamiltonian, both at the end of the ramp ($A = 1$), plotted as a function of the Landau-Zener parameter (main panel) and of the ramp time (inset). The dash-dotted line in the inset depicts $\tau^{-2}$ behavior. We show results for the same four lattice sizes as in Fig.~\ref{fig:lz-overlap}. (b) Excitation density $\rho_E$ in the time-evolved state plotted as a function of the ramp time. As in (a), we show results for the same four lattice sizes as in Fig.~\ref{fig:lz-overlap}. In addition, for the longest ramp times, we show results for $L=1024$. A fit of the $L=1024$ results to $1/\tau$ is shown as a dashed line. Inset: Discrete derivative $ \alpha=\Delta \ln  \rho_E / \Delta \ln (\tau/T) $ for the data shown in the main panel (for clarity, we do not show the results for $L=64$). Note the plateau at $\alpha=-1$, whose onset (the onset of the Kibble–Zurek regime) is marked by a vertical dotted line (at $\tau=50T$, also shown in the main panel). Another plateau ($\alpha=-2$, indicated by the arrow) shows the near-adiabatic regime. The latter moves to longer ramp times as the system size increases (the near-adiabatic regime is not visible for the two largest system sizes).}
\end{figure}

Figure~\ref{fig:lz-overlap} makes apparent that, with increasing system size, the maximal excitation is described by the Landau-Zener prediction for a wider range of Landau-Zener parameters. The collapse extends to smaller [Fig.~\ref{fig:lz-overlap}(a)] and larger (main panel in Fig.~\ref{fig:lz-overlap}) values of $\tau/(L^2T)$. We have found that the three regimes discussed for $P_E$ in Fig.~\ref{fig:lz-overlap} are also apparent in other quantities. For example, the Landau-Zener and near-adiabatic regimes can be seen in Fig.~\ref{fig:lz-overlap-full}(a) for the overlap between the entire time-evolved state $\ket{\Psi_\tau}$ and the lower Floquet band of the final Hamiltonian, $\ket{\Psi_F}$.

Another interesting aspect of crossing a topological phase transition in the thermodynamic limit is the fact that the excitation density $\rho_E = \sum_k P_{E,k} / L^2$ across the Brillouin zone should be governed by a Kibble–Zurek scaling $\rho_E \sim \tau^\alpha$~\cite{kibble_topology_1976, kibble_implications_1980, zurek_cosmological_1985, zurek_cosmological_1996, del_campo_universality_2014}. Integrating Eq.~\eqref{eq:LZ-excitation} over the Dirac cone, one finds that in the Landau-Zener regime $\alpha=-1$~\cite{privitera_quantum_2016, ulcakar_slow_2018, damski_simplest_2005}. In Fig.~\ref{fig:lz-overlap-full}(b) we plot $\rho_E$ vs $\tau/T$ for the same lattice sizes as in Fig.~\ref{fig:lz-overlap-full}(a). Those results show the existence of an expanding-with-system-size Landau-Zener regime with Kibble–Zurek scaling of $\rho_E$ (note that an increasing number of points collapse onto the $1/\tau$ line) that is preceded by a nearly size-independent fast-ramp regime and succeeded by a near-adiabatic one. The analysis in the inset in Fig.~\ref{fig:lz-overlap-full}(b), whose goal is to determine the power of a potential power-law decay for each pair of contiguous data points shown in the main panel, allows us to identify the onset of the Kibble–Zurek $1/\tau$ regime ($\tau \approx 50 T$, see vertical dotted lines). Note that $\tau \approx 50 T$ also marks the onset of the Landau-Zener regime in Fig.~\ref{fig:lz-overlap}(a). In Fig.~\ref{fig:lz-overlap-full}(b) and its inset one can see that in the near-adiabatic regime, which for the largest values of $\tau$ considered here is visible only in the three smallest system sizes shown, the excitation density is much smaller than in the Landau-Zener regime and scales as $\rho_E \sim \tau^{-2}$.

The existence of a Landau-Zener regime that, with increasing system size, extends to arbitrarily long ramp times allows us to describe analytically the scaling of the dynamical critical field strength $A^{\ast}_{L,\tau}$ at which the Chern number changes during the dynamics in finite systems. For that, we use that the time dependence of the probability to remain in the ground state (the ``adiabatic probability'') for the Landau-Zener problem can be expressed in terms of parabolic cylindrical functions $D_a(z)$~\cite{zener_non-adiabatic_1932},
\begin{equation}
  \label{para_cylin_func}
  P_A \left(\sqrt{\beta}t,\frac{V^2}{\beta}\right) = \sqrt{\frac{V^2}{2 \beta}} \mathe^{- \frac{\mathpi
      V^2}{8 \beta}}  \left[ D_{- \frac{ \mathi V^2 }{ 2 \beta } - 1} \left(
      \mathe^{ \mathi \frac{5 \mathpi}{4}} \sqrt{2 \beta} t \right) \right]^2.
\end{equation}
The dynamics in the Landau-Zener regime is then determined by both the Landau-Zener parameter and the dimensionless time $\sqrt{\beta} t$, as one could have concluded from a dimensional analysis. 

\begin{figure}[!t]
  \includegraphics[width=\columnwidth]{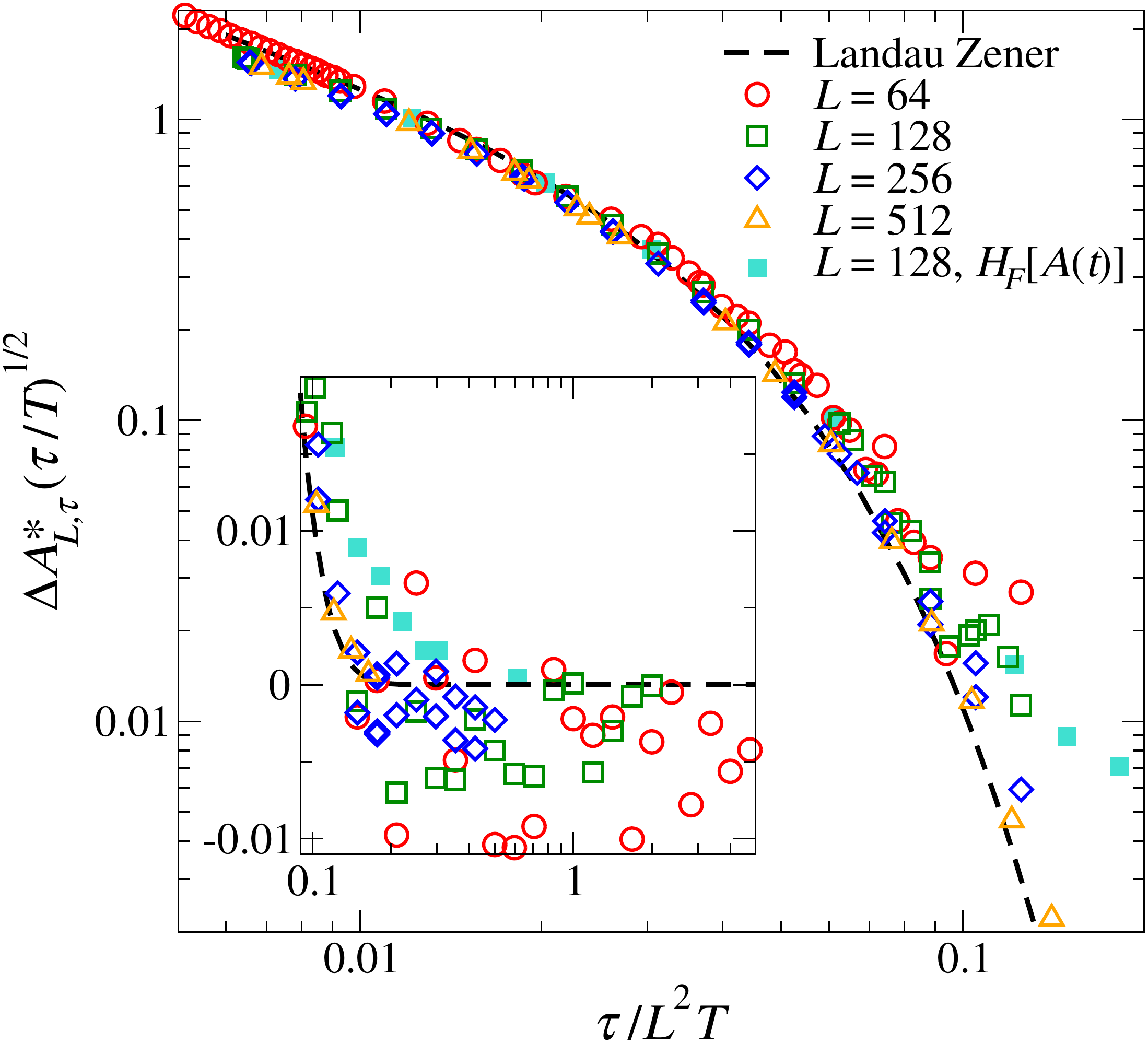}
  \caption{\label{fig:pbc-dyneaA-lz}
    Reduced deviation of the dynamical critical field $\Delta A^{\ast}_{L,\tau} ( \tau / T )^{1/2}$ plotted as a function of the Landau-Zener parameter for the same lattice sizes as in the previous figures. The dashed line shows the analytical Landau-Zener prediction using Eq.~\eqref{para_cylin_func}. For too-small values of the Landau-Zener parameter no dynamical transition occurs (this is why the data are cut at much larger values of the Landau-Zener parameter than in the previous figures). Inset: The same data plotted using a linear scale for the reduced deviation, whose absolute value converges to zero for all system sizes as $\tau \to \infty$.}
\end{figure}

For our ramps, the time at which $A(t) = A_L^{\ast}$ is the one that corresponds to $t=0$ in the Landau-Zener problem. We can then rewrite the dimensionless time at which the Chern number changes ($\sqrt{\beta} t^*$) in terms of the deviation between the dynamical ($A^{\ast}_{L,\tau}$) and Floquet ($A_L^{\ast}$) critical field strengths ($\Delta A^{\ast}_{L,\tau} = A^{\ast}_{L,\tau} - A_L^{\ast}$), namely, $\sqrt{\beta} t^* \sim \sqrt{\tau}  \Delta A^{\ast}_{L,\tau}$. In Fig.~\ref{fig:pbc-dyneaA-lz}, we plot $\Delta A^{\ast}_{L,\tau} (\tau/T)^{1/2}$ as a function of the Landau-Zener parameter. The results exhibit a clear data collapse for over a decade in the range of Landau-Zener parameters and, for the largest values of the Landau-Zener parameter shown, the collapse improves with increasing system size. We note that, on the other hand, for the smallest values of the Landau-Zener parameter shown the data are cut at much larger values of $\tau/(L^2T)$ than in Figs.~\ref{fig:lz-overlap} and~\ref{fig:lz-overlap-full}(a). This is because the Chern number either does not change or exhibits oscillatory behavior. In the presence of oscillations, we report $\Delta A^{\ast}_{L,\tau} (\tau/T)^{1/2}$ at the time the Chern number changes for the first time only if the state at the end of the ramp has a nonvanishing Chern number and remains so at long times under the evolution driven by the final Hamiltonian. Figure~\ref{fig:pbc-dyneaA-lz} shows that, with increasing system size, longer ramps are needed for the Chern number to change.\\

The data collapse in Fig.~\ref{fig:pbc-dyneaA-lz} indicates that $P_A \left(\sqrt{\beta}t^* , V^2/\beta\right)$ is constant at the time at which the dynamical critical field is crossed. Theoretically, we expect $P_A \left(\sqrt{\beta}t^* , V^2/\beta\right) = 1/2$ at the dynamical critical field. This can be visualized in the Bloch sphere picture for two-band systems \cite{ge_topological_2017,bernevig_topological_2013}. In a trivial initial state near the gap-crossing momentum $K'$, the states are within one hemisphere, leaving the other unoccupied. A threshold $P_A \left(\sqrt{\beta}t^* , V^2/\beta\right) = 1/2$ marks the entrance of the time-evolved state to the other hemisphere. Hence, we can compute $A^{\ast}_{L,\tau}$ analytically by setting $P_A \left(\sqrt{\beta}t^* , V^2/\beta\right) = 1/2$ in Eq.~\eqref{para_cylin_func}. We find that the range of Landau-Zener parameters for which $P_E$ in Eq.~\eqref{eq:LZ-excitation} is smaller than 1/2 (so that $P_A>1/2$) is essentially the same for which we report values of $\Delta A^{\ast}_{L,\tau} (\tau/T)^{1/2}$ in Fig.~\ref{fig:pbc-dyneaA-lz}. We focus on that range of Landau-Zener parameters when solving for $A^{\ast}_{L,\tau}$. Also, the fact that $P_A \left(\sqrt{\beta}t, V^2/\beta\right)$ vs $\sqrt{\beta}t$ is oscillatory for the smallest values of $\tau/(L^2T)$ shown in Fig.~\ref{fig:pbc-dyneaA-lz} allows us to understand the oscillatory behavior observed in the Chern number in that regime. In the presence of oscillations that cross $1/2$, $\tau^*$ is taken to be the first time at which $P_A \left( \sqrt{\beta}t, \frac{V^2}{\beta}\right)$ crosses 1/2.

Figure~\ref{fig:pbc-dyneaA-lz} shows that the theoretical curve extracted this way agrees well with the numerical results when the latter exhibit data collapse. When the data departs from the analytic prediction for large values of the Landau-Zener parameter, the dynamical critical field can be lower than its equilibrium counterpart. This is shown in the inset of Fig.~\ref{fig:pbc-dyneaA-lz} and is the result of micromotion, namely, of the fact that the system can closely follow the oscillatory electric field within each driving period~\cite{shirley_solution_1965, sambe_steady_1973, bukov_universal_2015}. To verify that micromotion is indeed the cause, we carried out the evolution of a system with $L=128$ using the Floquet Hamiltonian corresponding to the instantaneous field strength $H_F [ A ( t )]$. The results are also reported in Fig.~\ref{fig:pbc-dyneaA-lz} with solid symbols (denoted as $L=128,\ H_F[ A ( t )]$). In that case, $\Delta A^{\ast}_{L,\tau}$ remains positive as it approaches zero with increasing Landau-Zener parameter.

\subsection{Cylinder geometry}

Here we carry out the same analysis for cylinders. As mentioned in Sec.~\ref{sec:equilibrium}, in order to minimize finite-size effects, we focus on cylinders with $L=3n+2$.

\begin{figure}[!t]
  \includegraphics[width=\columnwidth]{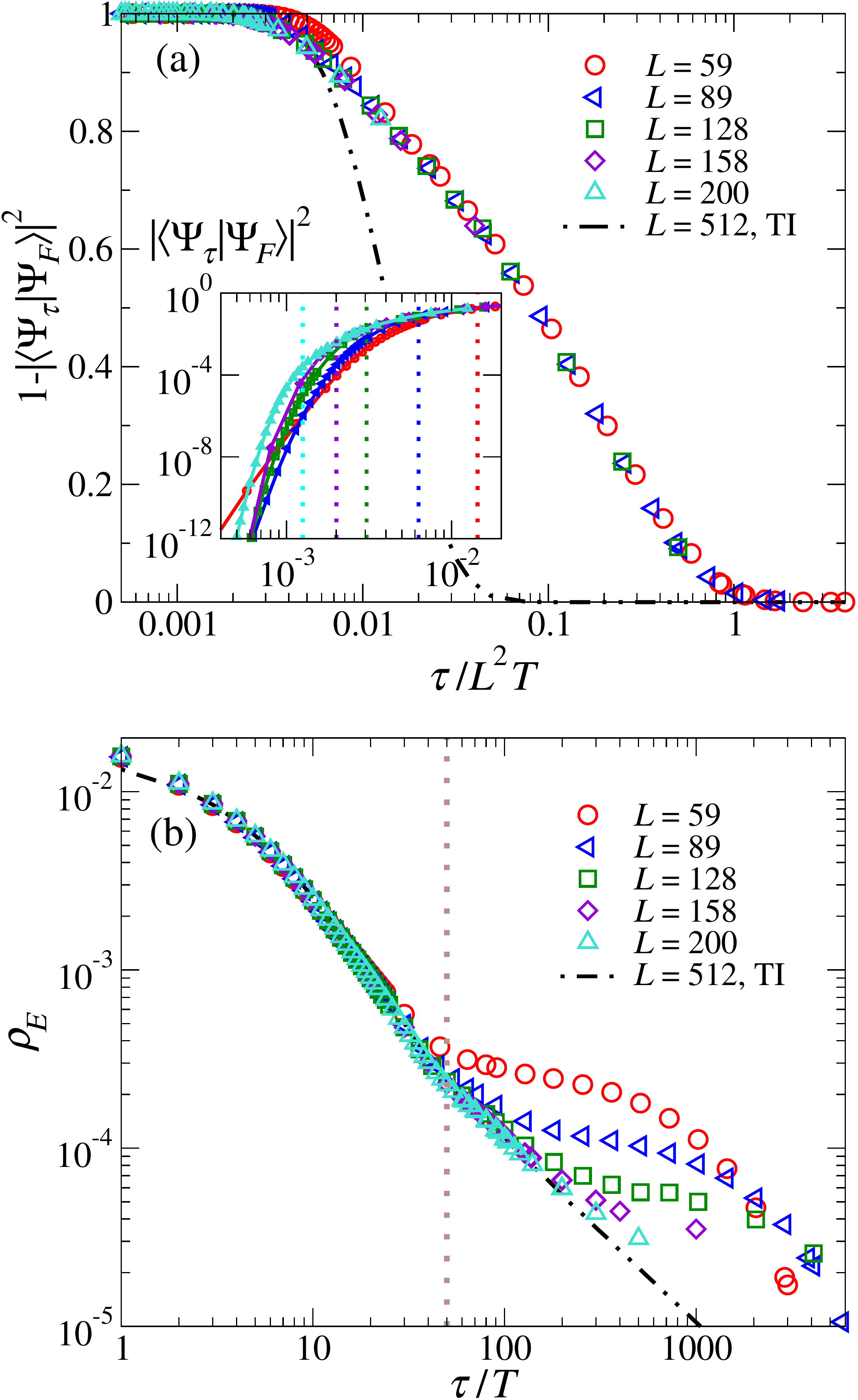}
  \caption{\label{fig:cylinder-overlap}
    (a) $ 1 - | \langle \Psi_\tau | \Psi_{F} \rangle |^2 $, where $\ket{\Psi_\tau}$ is the time-evolved state and $| \Psi_{F} \rangle$ is the appropriate reference state of the Floquet Hamiltonian (see explanation in the text), both at the end of the ramp ($A = 1$), plotted as a function of the Landau-Zener parameter in the cylinder geometry. For comparison, the dash-dotted line ($L=512$, TI) shows the results obtained for the translationally invariant system with $L=512$ reported in Fig.~\ref{fig:lz-overlap-full}(a). Inset: Overlap $| \langle \Psi_\tau | \Psi_{F} \rangle |^2$ vs the Landau-Zener parameter in the fast-ramp regime. (b) Excitation density $\rho_E$ in the time-evolved state plotted as a function of the ramp time. For comparison, the dash-dotted line ($L=512$, TI) shows the results obtained for the translationally invariant system with $L=512$ reported in Fig.~\ref{fig:lz-overlap-full}(b). The vertical dotted lines in the inset in (a) and in (b) mark $\tau=50T$, the ramp time at which the onset of the Landau-Zener regime was identified in Fig.~\ref{fig:lz-overlap-full} for incommensurate translationally invariant systems.}
\end{figure}

Quantifying how much, e.g., in terms of overlaps and the amount of excitations, cylinders are affected by crossing a topological transition is complicated by the presence of edge states. Out of equilibrium, the far-separated topological edges are effectively two disconnected one-dimensional systems with possibly different chemical potentials. Hence, simply taking the ground state of the final Floquet Hamiltonian to be the reference state for adiabatic dynamics may not yield meaningful results~\cite{privitera_quantum_2016, liou_quench_2018, ulcakar_slow_2018}. As discussed by Privitera and Santoro~\cite{privitera_quantum_2016}, in the thermodynamic limit edge states of the instantaneous Floquet Hamiltonian (with vanishing coupling to any other state, including those at the opposite edge) can move around in the bulk gap during the dynamics. As a result, when the crossing point of the dispersion of opposite edge states shifts, occupied states end up having a higher energy than unoccupied ones even though no true excitation has occurred. Those occupied states need to be identified and included in the reference state used to quantify excitations. Failing to do this can result, e.g., in a vanishing overlap between the reference state and a near-adiabatic time-evolved one.

For the Floquet Hamiltonian in cylinders, our reference state for adiabatic dynamics $\ket{\Psi_F}$ contains the lower bulk band and all edge states on both sides of the cylinder. Consequently, we do not resolve the contributions of the bulk and the edges independently. For the zigzag edges considered here, the edge modes extend between the two Dirac points, $2\pi/3 \le k a \le 4\pi/3$, where $a = \sqrt{3} d$ is the unit cell spacing~\cite{bernevig_topological_2013}. Hence, we take the lowest $L+1$ states for each $k$ within this range and $L$ states elsewhere to form the reference basis. We then compute the occupation matrices $\braket{\psi_\tau^i(k)}{\psi_F^j(k)}$, with $i = 1, \ldots, L$ and $j = 1, \ldots, L(+1)$ outside (inside) the edge-state momentum range. The $L$ singular values of this matrix are the square roots of occupation numbers. They are used to compute the overlap (the product of the square of the singular values) and the excitation density (the mean of the square of the singular values). One can verify the correctness of the reference state looking at the spectrum of these singular values. It should be continuous in $k$ and change sharply when the reference state filling number is one too many.

The overlap obtained this way is shown in Fig.~\ref{fig:cylinder-overlap}(a). It exhibits collapse when plotted against the Landau-Zener parameter, for $\tau \gtrsim 50T$ (see inset), even though the falloff is not as fast as in the incommensurate translationally invariant geometry (see the dash-dotted line for the results for $L=512$). The corresponding excitation density in Fig.~\ref{fig:cylinder-overlap}(b) obeys the Kibble-Zurek prediction. Notice the collapse with increasing system size to the results obtained for the incommensurate translationally invariant geometry with $L=512$ (dash-dotted line).

\begin{figure}[!t]
	\includegraphics[width=\columnwidth]{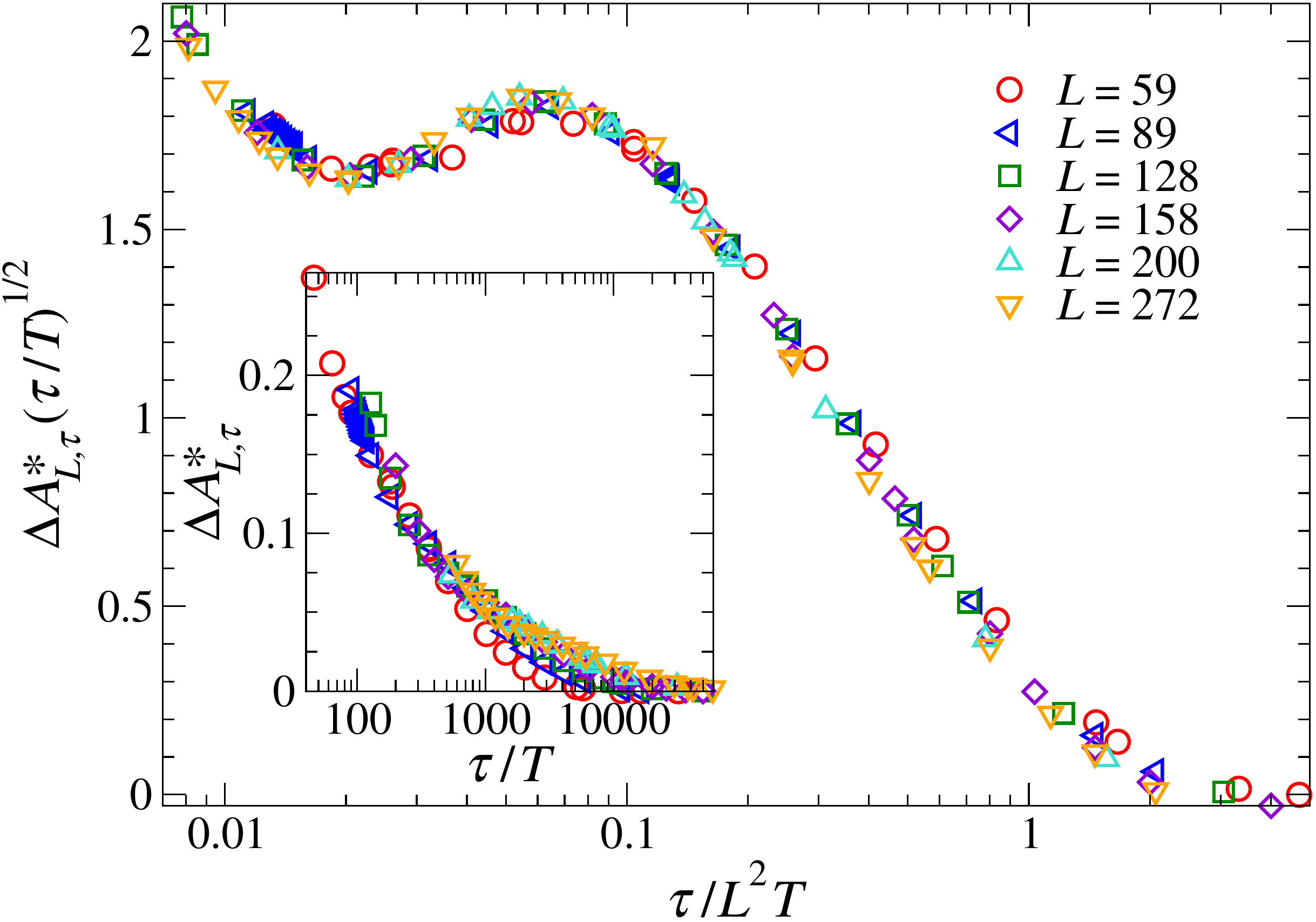}
	\caption{\label{fig:cylinder-dyneaA-lz}
	    Reduced deviation of the dynamical critical field $\Delta A^{\ast}_{L,\tau} ( \tau / T )^{1/2}$ plotted as a function of the Landau-Zener parameter in the cylinder geometry. For too-small values of the Landau-Zener parameter no dynamical transition occurs (this is why the data are cut at larger values of the Landau-Zener parameter than in Fig.~\ref{fig:cylinder-overlap}). For the slowest ramps, $\Delta A^{\ast}_{L,\tau}$ sometimes becomes negative (${\sim}{-10^{-5}}$). Inset: The unscaled $\Delta A^{\ast}_{L,\tau}$ vs $\tau/T$, which decreases (mostly) monotonically. Using the approach discussed in Appendix~\ref{sec:efficient-bott}, we are able to compute $A^{\ast}_{L,\tau}$ for larger lattices and longer ramp times than for the overlaps and excitation densities reported in Fig.~\ref{fig:cylinder-overlap}.}
\end{figure}

Given the results reported in Fig.~\ref{fig:cylinder-overlap}, it is natural to expect that the critical field for the dynamical topological phase transition in the cylinder geometry exhibits a Landau-Zener regime that extends to larger values of $\tau$ as the system size increases. Indeed, the behavior of the reduced dynamical critical field $\sqrt{\tau} \Delta A_{L,\tau}^\ast$ in Fig.~\ref{fig:cylinder-dyneaA-lz} displays the advanced scaling collapse and convergence with increasing cylinder size. Interestingly, and in contrast to the results for incommensurate translationally invariant systems, for the first decade of Landau-Zener parameters shown $\sqrt{\tau} \Delta A_{L,\tau}^\ast$ is not monotonic. However, we should stress that, as shown in the inset in Fig.~\ref{fig:cylinder-overlap}, the dynamical critical field itself is mostly monotonic, up to small oscillations similar to the ones observed in the translationally invariant case. For very slow ramps micromotion again sometimes results in negative values of $\Delta A_{L,\tau}^\ast$.

\subsection{Patch geometry}

The patch geometry has the largest finite-size effects, as seen in the finite-size scaling analysis of the critical field for the topological transition in the Floquet Hamiltonian (see Fig.~\ref{fig:eq-trans}). Also, in the topological phase, the distinction between edge and bulk states is ambiguous. In fact, there are (sometimes discontinuous) ranges of fillings within the bulk gap for which the Bott index is nonzero~\cite{dalessio_dynamical_2015}. These issues make it difficult to define a reference state $\ket{\Psi_F}$, and to carry out analyses such as the ones reported in Fig.~\ref{fig:cylinder-overlap} for cylinders. Hence, for patches we focus on the scaling of the dynamical critical field.

\begin{figure}[!b]
  \includegraphics[width=\columnwidth]{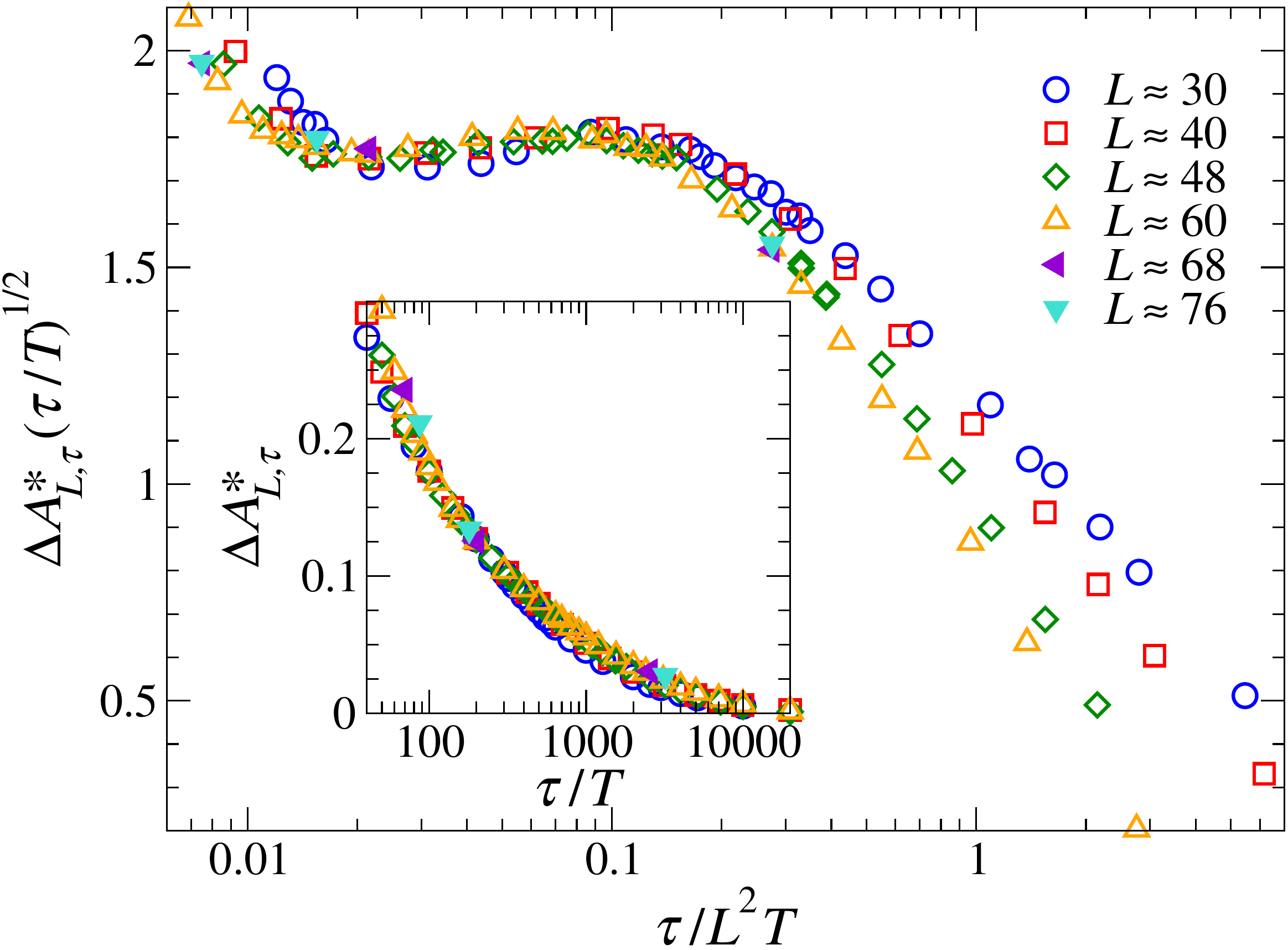}
  \caption{\label{fig:opatch-dyneaA-lz}
	    Reduced deviation of the dynamical critical field $\Delta A^{\ast}_{L,\tau} ( \tau / T )^{1/2}$ plotted as a function of the Landau-Zener parameter in the patch geometry. For too-small values of the Landau-Zener parameter no dynamical transition occurs. Inset: The unscaled $\Delta A^{\ast}_{L,\tau}$ vs $\tau/T$, which decreases (mostly) monotonically. For $L\approx68$ and 76 we only show results for three ramp speeds. They help gauging convergence with increasing the size of the patch. Qualitatively, the behavior of the dynamical critical field is similar to that in the cylinder geometry reported in Fig.~\ref{fig:cylinder-dyneaA-lz}. For the longest ramps and the largest system sizes, the Bott index is computed with the method in Appendix~\ref{sec:efficient-bott}.}
\end{figure}

In Fig.~\ref{fig:opatch-dyneaA-lz} we show the reduced deviation of the dynamical critical field plotted as a function of the Landau-Zener parameter. The results are qualitatively similar to the ones in Fig.~\ref{fig:cylinder-dyneaA-lz} for cylinders. As in the cylinder geometry, we find that $\sqrt{\tau} \Delta A_{L,\tau}^\ast$ exhibits nonmonotonic behavior, while the dynamical critical field itself is mostly monotonic (see inset in Fig.~\ref{fig:opatch-dyneaA-lz}). Also, the data are consistent with a collapse that improves and extends to larger values of the Landau-Zener parameter with increasing patch size. We note that, in Fig.~\ref{fig:opatch-dyneaA-lz}, only patches with $L\approx30$, 40, 48, and 60 were systematically studied for many ramp times $\tau$. Due to their high computational cost, we report results for $L\approx68$ and 76 only for three ramp times. They are consistent with the trend seen for smaller patches.

\section{\label{sec:hall}Hall Responses}

Our results in Sec.~\ref{sec:dynamics} show that, when ramping up the strength of the field of the drive in increasingly large systems (no matter the boundary conditions), topological indices such as the Chern number and the Bott index fail to indicate the crossing of a topological transition. Here we explore what happens with the more experimentally relevant Hall responses. We focus on translationally invariant systems as, based on our results in Sec.~\ref{sec:dynamics}, we expect cylinder and patch geometries to behave similarly for large systems sizes. 

For a static insulating state, the Hall conductivity given by linear response depends exclusively on the Berry curvature of the occupied band \cite{thouless_quantized_1982, xiao_berry_2010}. Out of equilibrium, the Hall conductivity depends both on the time-evolving state and the Hamiltonian~\cite{caio_hall_2016, xu_scheme_2019, ulcakar_slow_2018}. In this section we report results for the Hall response of the ground state of the Floquet Hamiltonian, as well as of the out-of-equilibrium state generated by ramping up the strength of the driving field, in translationally invariant systems.

We probe those states via turning on a weak constant electric field $E_x$ and measuring the transverse displacement. The displacement per unit cell along the primitive lattice vectors $\mathbf{a}_1$ and $\mathbf{a}_2$ is computed evaluating~\cite{resta_quantum-mechanical_1998}
\begin{equation}
 \label{eq:displacement} Y_{i}(t) = \frac{a}{2 \pi L} \sum_\mathbf{k} \tmop{Im} \ln \braket{u_\mathbf{k}(t)}{u_{\mathbf{k}+\mathrm{d}\mathbf{k}_{i}}(t)}, \quad i=1,2,
\end{equation}
where $u_\mathbf{k}$ is the periodic part of the Bloch state at momentum $\mathbf{k}$, and $\mathrm{d}\mathbf{k}_{1,2}$ is the displacement between neighboring $\mathbf{k}$ points along the two reciprocal vectors. From the stroboscopic transverse displacement $Y=(Y_1 - Y_2)/2$ over time, we obtain the time-averaged Hall conductivity computing
\begin{equation}
\label{eq:conductivity} \bar{\sigma}_{xy}(t)=\frac{2\pi}{S} \frac{Y(t)}{E_x t}
\end{equation}
in units of $e^2/h$, where $S$ is the area of a unit cell. Each term in the sum in Eq.~\eqref{eq:displacement} needs to be monitored over time to keep them on a continuous Riemann sheet. We favor the numerical evaluation of the displacement over the current because within each cycle the current produced by the Floquet drive and nonequilibrium state overwhelms the response to the applied field.

\subsection{Floquet ground state}

\begin{figure}[!t]
  \includegraphics[width=1\columnwidth]{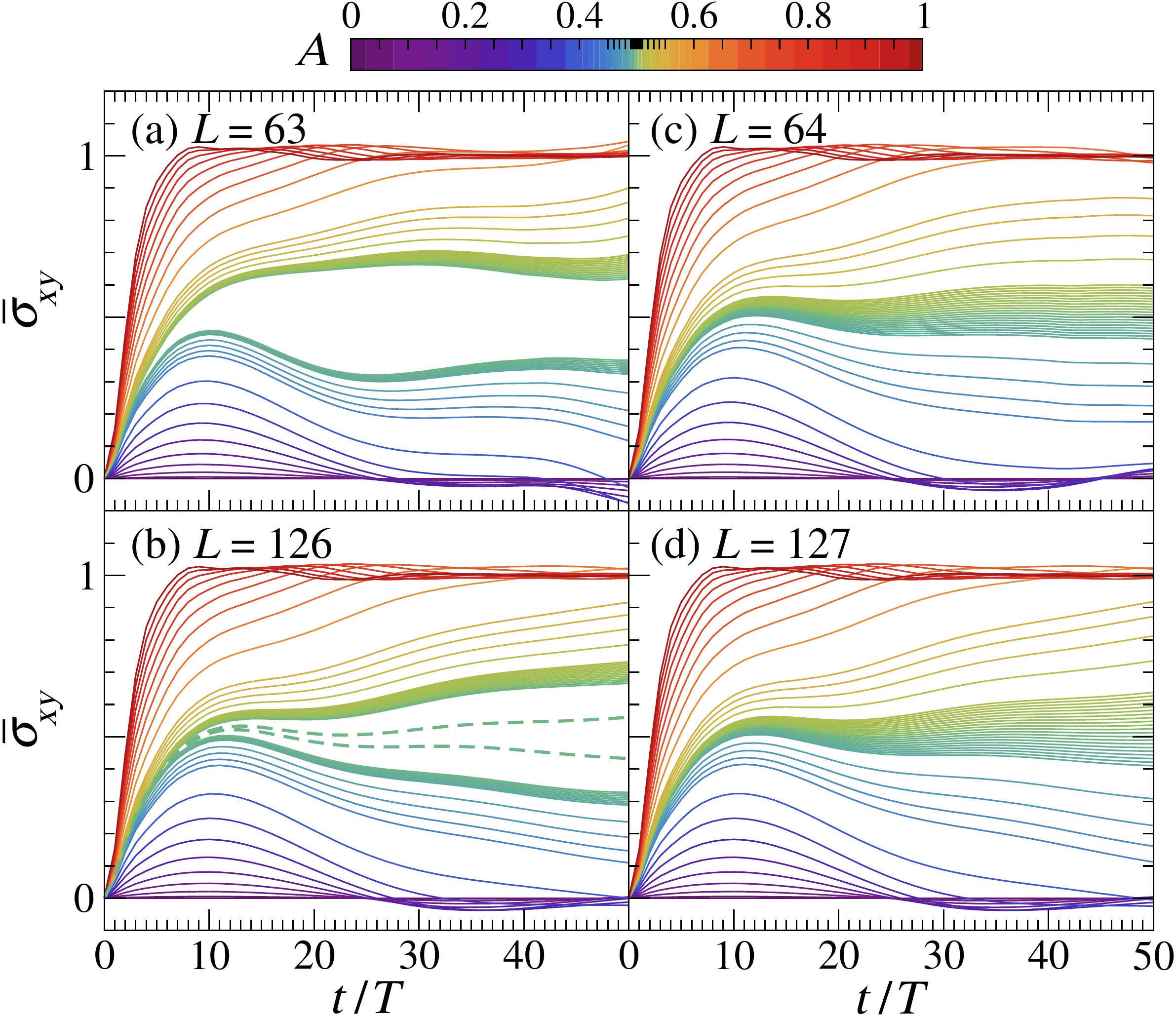}
  \caption{\label{fig:floquet-jy}
    Time-averaged Hall response $\bar{\sigma}_{xy}$ [see Eq.~\eqref{eq:conductivity}] of the ground state of the Floquet Hamiltonian in translationally invariant systems, in units of $e^2/h$. We show results for commensurate lattices, (a) $L=63$ and (b) $L=126$, as well as for incommensurate ones, (c) $L=64$ and (d) $L=127$. The strength of the Floquet driving field is taken between 0 and 1 and is color coded in the lines according to the legend at the top. The ticks in the color bar mark the field strengths $A$ for which results are reported (notice the finer $0.001$ resolution grid about the critical field for the topological transition). The dashed lines in (b) show the response for $A=0.497$ and 0.498 in a lattice with $L=300$. Those are the fields closest to the critical one for which results are reported in all panels.}
\end{figure}

In Fig.~\ref{fig:floquet-jy}, we plot the time-averaged Hall conductivity during the first 50 periods of the drive after turning on a weak constant electric field $E_x = 10^{-5}$. The initial state is taken to be the ground state of the Floquet Hamiltonian for driving field strengths between zero and 1 (see legend at the top of Fig.~\ref{fig:floquet-jy}). We report results for commensurate lattices $L=63$ [Fig.~\ref{fig:floquet-jy}(a)] and $L=126$ [Fig.~\ref{fig:floquet-jy}(b)], as well as for incommensurate ones $L=64$ [Fig.~\ref{fig:floquet-jy}(c)] and $L=127$ [Fig.~\ref{fig:floquet-jy}(d)].

The results in Fig.~\ref{fig:floquet-jy} show that, deep in the trivial ($A\rightarrow0$) and topological ($A\rightarrow1$) phases, the time-averaged Hall conductivity rapidly approaches its quantized zero and 1 values, respectively. As the ground state of the system is taken closer to the transition point ($A^{\ast} \approx 0.498$ in the thermodynamic limit) it takes much longer for the time-averaged Hall conductivity to become zero (if $A<A^{\ast}$) or 1 if (if $A>A^{\ast}$). The presence of the gap-closing momentum in finite commensurate systems [Figs.~\ref{fig:floquet-jy}(a) and~\ref{fig:floquet-jy}(b)] produces a ``jump'' between the dynamical responses for $A$ smaller and greater than $A^{\ast}$, which is absent in the incommensurate systems [Figs.~\ref{fig:floquet-jy}(c) and~\ref{fig:floquet-jy}(d)]. The jump in the former is the result of the large Berry curvature concentrated at the gap-closing momentum about the topological phase transition. 

We expect that, with increasing system size, the Hall responses of commensurate and incommensurate systems become similar to each other, with no jump (as a result of the smooth interpolation of the Berry curvature) and a wide range of values of $\bar{\sigma}_{xy}$ about the topological transition at the longest times shown (because of the increasing resolution of momenta about the gap-closing point). Our results in Fig.~\ref{fig:floquet-jy} are consistent with those expectations. Notice that in commensurate systems the magnitude of the jump decreases with increasing system size. This is apparent when comparing the results for $A=0.497$ and 0.498 in the lattice with $L=63$ [Fig.~\ref{fig:floquet-jy}(a)] and in the lattices with $L=126$ and 300 [Fig.~\ref{fig:floquet-jy}(b)]. In incommensurate systems, on the other hand, the response near the phase boundary fans out with increasing system size. In Fig.~\ref{fig:floquet-jy}, the average response has mostly acquired a quantized value after 50 driving periods, except for fields close to the critical one ($0.45 \le A \le 0.55$).

\subsection{Time-evolved state}

\begin{figure}[!b]
  \includegraphics[width=1\columnwidth]{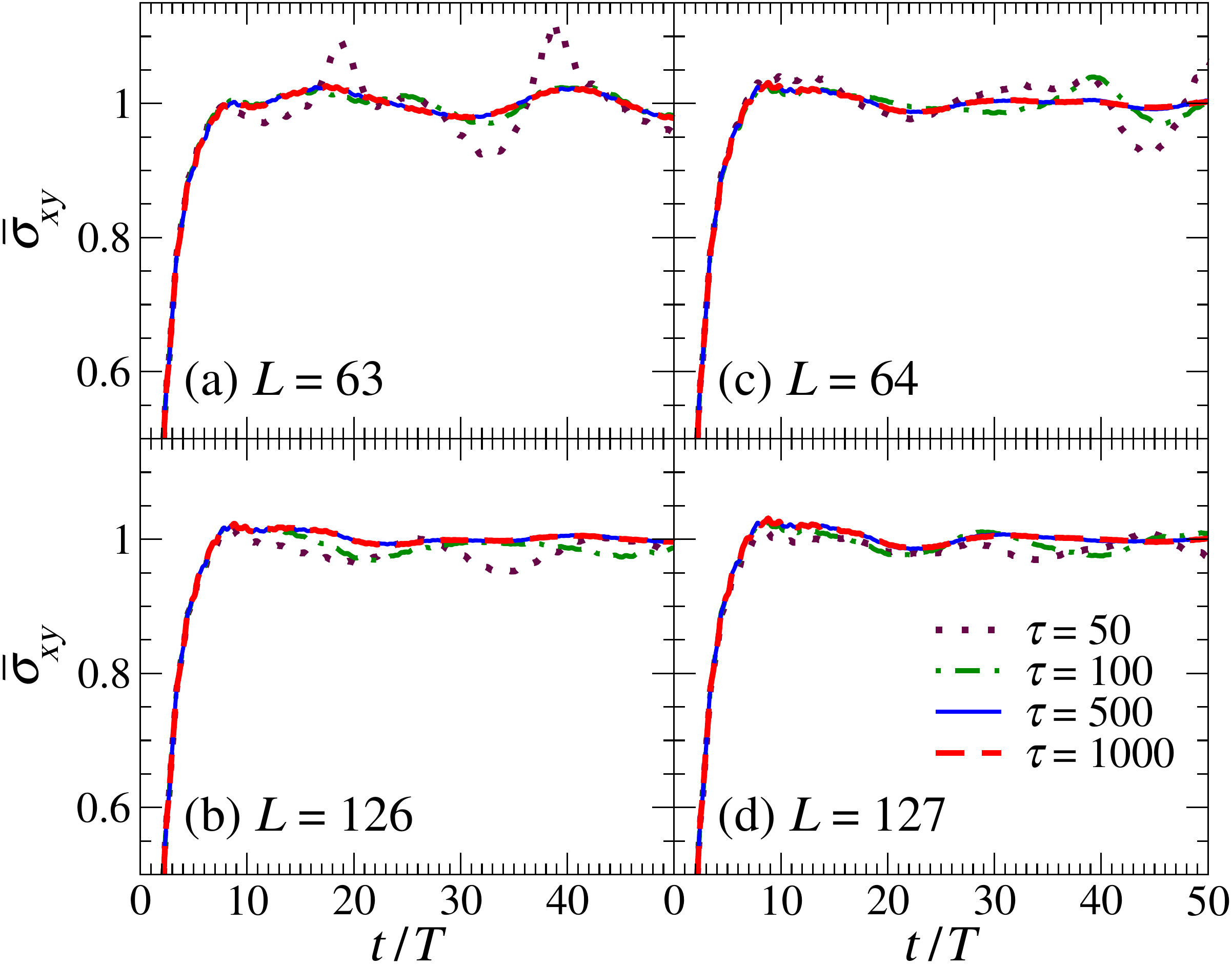}
  \caption{\label{fig:ramp-jy}
      Time-averaged Hall response $\bar{\sigma}_{xy}$ [see Eq.~\eqref{eq:conductivity}] of the time-evolved state prepared under different ramp times $\tau$ (the strength of the field is always increased from $A=0$ to 1) in translationally invariant systems, in units of $e^2/h$. After the ramp, we evolve the systems for long times (at fixed $A=1$) to probe the state after equilibration (see text). We show results for commensurate lattices, (a) $L=63$ and (b) $L=126$, as well as for incommensurate ones, (c) $L=64$ and (d) $L=127$, as in Fig.~\ref{fig:floquet-jy}. Independently of whether the Chern number changes or not, the Hall conductivity converges to the equilibrium quantized value with increasing ramp time and system size.}
\end{figure}

Before turning on $E_x$ to study the response of the time-evolved states obtained at the end of the ramp, we let those states equilibrate by time-evolving them for long times with the final Hamiltonian (with $A=1$). The equilibration simulations are run for between $10^3$ and $10^4$ periods under the final Hamiltonian. They allow the currents that accumulate during the ramp to relax. Such currents can be generated, e.g., by the higher strength of the driving field during the second half of the driving cycles. After equilibration, we measure the displacement after turning on a weak constant electric field $E_x = 10^{-5}$ and also without the field. We define the net displacement as the difference between the two. 

The Hall response extracted using the net displacement is shown in Fig.~\ref{fig:ramp-jy}. Regardless of whether the lattices considered are commensurate or not, and of whether the Chern number changes or not during the ramp (it can only change in incommensurate lattices), the time-evolved states exhibit nontrivial Hall responses that converge to the equilibrium value with increasing ramp time $\tau$ and system size $L$. This demonstrates that when crossing a Floquet topological transition under driven unitary dynamics it is possible to create states with nearly quantized Hall response, despite the fact that those states may have a trivial Chern number. The oscillatory behavior observed in Fig.~\ref{fig:ramp-jy} is a consequence of the coherent superposition of states in the lower and upper Floquet band \cite{xu_scheme_2019, ulcakar_slow_2018}. The magnitude of the oscillations can be seen to decrease with increasing ramp time and system size. Beyond ramp times $\tau \simeq 500$, the change in Hall response is not discernible in Fig.~\ref{fig:ramp-jy}.

\section{\label{sec:summary}Summary and discussion}

We have identified and explored the three dynamical regimes (fast-ramp, Landau-Zener, and near-adiabatic regimes) that occur in various observables when ramping up the strength of a driving field across a topological phase transition in finite incommensurate periodic lattices as well as in cylinder and patch geometries. 

For observables such as the maximal excitation, the overlap between the time-evolved state and the appropriate lowest-energy Floquet reference state, and the average excitation energy, whenever they can be meaningfully defined, we find that the fast-ramp regime is nearly system-size independent. This regime is followed (for longer ramp times) by a Landau-Zener one whose extent in ramp times increases with increasing system size. In the Landau-Zener regime we showed that the excitation density is governed by a Kibble–Zurek scaling $\rho_E \sim \tau^{-1}$. On the other hand, for very slow ramps, we showed that there is a near-adiabatic regime in which the magnitude of the observables is governed by the square of the ramp speed.

In the fast-ramp regime, and in the initial faster ramp part of the Landau-Zener regime identified via the previously mentioned observables, we find that with increasing system size the topological indices do not change (or oscillate) when crossing the topological transition. Consistent with the fact that the topological indices are invariant in the thermodynamic limit, we have shown that as the size of the lattices increases longer ramp times are needed for the topological indices to change. When they do change, we observe a robust regime in which the dynamical critical field scales with the Landau-Zener parameter. That regime extends to larger values of the Landau-Zener parameter with increasing system size. 

Finally, we showed that the dc Hall response allows one to identify the dynamical crossing of a Floquet topological phase transition independently of the behavior of the topological indices. Specifically, we showed that the states created under driven unitary dynamics after crossing a Floquet topological transition have nearly quantized Hall response when the strength of the driving field is increased using slow enough ramps. In contrast to the Chern number and the Bott index, the ramp times needed to observe a nearly quantized Hall response are $O(1)$, namely, they do not diverge with increasing system size.

\begin{acknowledgments}
  We acknowledge support from the National Science Foundation under Grant No.~PHY-2012145. The computations were done in the Roar supercomputer of the Institute for Computational and Data Sciences (ICDS) at Penn State.
\end{acknowledgments}

\appendix

\section{\label{sec:cylinder-bott}Bott index in the cylinder geometry and its relation to the local Chern marker}

Here we provide a derivation of the Bott index expression in Eq.~\eqref{eq:bott-cylinder}. We note that the original Bott index formula, albeit computationally more costly, can be used in cylinders. The Bott index is
\begin{equation}\label{eq:bott-def}
C_b (\hat{P}) = \frac{1}{2 \mathpi} \tmop{Im} \tmop{Tr} \ln (\tilde{V} \tilde{U} \tilde{V}^{\dag}  \tilde{U}^{\dag}),
\end{equation}
where $\tilde{U}$ and $\tilde{V}$ are the operators $\hat{U}:= \mathe^{2 \mathpi \mathi \hat{x} / L_x}$ and $ V:=\mathe^{2 \mathpi \mathi \hat{y} / L_y}$, respectively, projected onto the occupied states. In cylinder geometries there is translational invariance in one direction ($x$ in our case), so the $\hat{U}$ operator can be written as
\begin{eqnarray}
\hat{U} & = & \sum_{k_x, \alpha} \left| k_x + \frac{2 \mathpi}{L_x}, \alpha \middle\rangle  \middle\langle k_x, \alpha \right| \nonumber\\*
 & = & \sum_{k_x} \left| k_x + \frac{2\mathpi}{L_x} \middle\rangle  \middle\langle k_x \right| \otimes I_\alpha,
\end{eqnarray}
where $\alpha$ are the other degrees of freedom, such as the cell index in the $y$ direction and sublattice sites, and $I_\alpha$ is the identity operator in these degrees of freedom. Eigenstates of the Hamiltonian are Bloch states along $x$. As a result, the projection operator has the form
\begin{eqnarray}
\hat{P} & = & \sum_{k_x n} \ket{\psi_{k_x n}} \bra{\psi_{k_x n}} \nonumber\\*
 & = & \sum_{k_x n} \mathe^{\mathi k_x \hat{x}} \ket{u_{k_x n}} \bra{u_{k_x n}} \mathe^{- \mathi k_x \hat{x}} \nonumber\\*
 & = & \sum_{k_x} \mathe^{\mathi k_x \hat{x}} \hat{P}_{k_x} \mathe^{- \mathi k_x \hat{x}},
\end{eqnarray}
where $\{\ket{\psi_{k_x n}}\}$ are the Bloch eigenstates of band $n$, $\ket{u_{k_x n}}$ is the lattice periodic part of the corresponding Bloch state, and $\hat{P}_{k_x} \assign \sum_n \ket{u_{k_x n}} \bra{u_{k_x n}}$ is the projection onto occupied bands at $k_x$. Hence
\begin{equation}\label{eq:tildeu}
\tilde{U} = \hat{P}  \hat{U}  \hat{P} = \sum_{k_x} \hat{P}_{k_{x} + \frac{2 \mathpi}{L_x}} \hat{P}_{k_x} .
\end{equation}
Plugging $\tilde{U}$ from Eq.~\eqref{eq:tildeu} in Eq.~\eqref{eq:bott-def} one obtains Eq.~\eqref{eq:bott-cylinder}.

We note that another topological index of choice for systems with open boundary conditions is the local Chern marker. It can be used to probe the Chern number locally in the real space. It is defined as
\begin{equation}
    c(\mathbf{r}) = \frac{4\pi}{S} \tmop{Im} \sum_\alpha \bra{\mathbf{r}_\alpha} \hat{P} \hat{x} \hat{P} \hat{y} \hat{P} \ket{\mathbf{r}_\alpha},
\end{equation}
where $S$ the unit cell area, and the sum takes place over the orbitals $\alpha$ in the unit cell at $\mathbf{r}$. When averaged over an entire patch, the local Chern marker yields zero, reflecting the fact that an open boundary system is topologically trivial as a whole~\cite{bianco_mapping_2011}. This comes from the large negative values that the local Chern marker takes at the edges. The physical picture is that the cyclotronic bulk and the skipping orbits at the edge have opposite chiralities. The Bott index, on the other hand, eliminates the edge contribution by gluing opposite edges together and allowing their chiralities to cancel. Therefore, the Bott index can be seen as the bulk average of the local Chern marker.

\section{\label{sec:efficient-bott}Efficient calculation of the dynamics of topological indices}

To detect a topological phase change for long enough ramps in translationally invariant systems, it is sufficient to track the time evolution of states near the gap-closing momenta. It is not necessary to carry out the time evolution of the entire Brillouin zone. The same turns out to be true for the other two geometries considered in this work.

In the cylinder geometry, we also find that only states near the gap-closing momenta need to be time evolved for long enough ramps. One can further notice that, in the initial Hamiltonian, there are already edge states at the top of the valence band. They also end up with the highest overlap with the topological edge states. For very long ramps only the top two states of the initial valence band at each momentum are needed to capture the jump in Berry curvature. Compared to the full band calculations, the error introduced in $A^{\ast}_{L,\tau}$ and $\Delta A^{\ast}_{L,\tau} ( \tau / T )^{1/2}$ in slow ramps is smaller than the size of the symbols in Fig.~\ref{fig:cylinder-dyneaA-lz}.

In the patch geometry, the calculations for very long ramps also benefit from the presence of initial edge states. Here the Bott index of a valence band can be similarly computed starting from the top of the valence band as follows. Denote a generic projection matrix by $\hat{P}(l,h) = \sum_{i=l}^h \ket{i} \bra{i}$, where $\ket{i}$ is the $i$th eigenstate. In a typical Bott index calculation the indices $[l,h]$ enclose the valence band and possibly topological edge states. Instead, one can let $l$ start somewhere in the bulk and still obtain a well-behaved Bott index across in the phase diagram. Thanks to the initial edge states, we can also track the dynamics of the Bott index in this way. The resulting error in $A^\ast_{L,\tau}$ and $\Delta A^{\ast}_{L,\tau} ( \tau / T )^{1/2}$ is again indiscernible in our results in Fig.~\ref{fig:opatch-dyneaA-lz}.

\bibliography{Biblio-Database,Rigol}

\end{document}